\documentclass[%
 reprint,
 superscriptaddress,
 amsmath,amssymb,
 aps,
 pra,
 floatfix,
]{revtex4-2}

\newcommand{\bra}[1]{\langle #1\rvert}
\newcommand{\ket}[1]{\lvert #1\rangle}

\DeclareMathOperator{\Tr}{Tr}

\usepackage{graphicx}
\usepackage{dcolumn}
\usepackage{bm}
\usepackage{hyperref}

\usepackage{tikz}
\usepackage{dsfont}
\usetikzlibrary{calc}

\usepackage{qcircuit}

\usepackage{amsmath}
\usepackage{upgreek}
\usepackage{physics}
\usepackage{isomath}
\usepackage{placeins}

\begin{document}


\title{Beyond one-axis twisting: Simultaneous spin-momentum squeezing}

\author{John Drew Wilson}
\thanks{Corresponding author: J. D. W.; John.Wilson-6@colorado.edu}
\affiliation{JILA and Department of Physics, University of Colorado, 440 UCB, Boulder, CO 80309, USA}
\author{Simon B. J\"ager}
\affiliation{JILA and Department of Physics, University of Colorado, 440 UCB, Boulder, CO 80309, USA}
\affiliation{Physics Department and Research Center OPTIMAS, Technische Universit\"at Kaiserslautern, D-67663, Kaiserslautern, Germany}
\author{Jarrod  T.  Reilly}
\affiliation{JILA and Department of Physics, University of Colorado, 440 UCB, Boulder, CO 80309, USA}
\author{Athreya Shankar}
\affiliation{Institute for Theoretical Physics, University of Innsbruck, Innsbruck, Austria}
\affiliation{Institute for Quantum Optics and Quantum Information of the Austrian Academy of Sciences, Innsbruck, Austria}
\author{Maria Luisa Chiofalo}
\affiliation{%
 Dipartimento di Fisica ``Enrico Fermi'', Università di Pisa, and INFN \\
 Largo Bruno Pontecorvo, 3 I-56127 Pisa (Italy) 
}%
\author{Murray J. Holland}
\affiliation{JILA and Department of Physics, University of Colorado, 440 UCB, Boulder, CO 80309, USA}
\affiliation{%
 Dipartimento di Fisica ``Enrico Fermi'', Università di Pisa, and INFN \\
 Largo Bruno Pontecorvo, 3 I-56127 Pisa (Italy) 
}%


\date{\today}

\begin{abstract}
The creation and manipulation of quantum entanglement is central to improving precision measurements. A principal method of generating entanglement for use in atom interferometry is the process of spin squeezing whereupon the states become more sensitive to $SU(2)$ rotations. One possibility to generate this entanglement is provided by one-axis twisting (OAT), where a many-particle entangled state of one degree of freedom is generated by a non-linear Hamiltonian. We introduce a novel method which goes beyond OAT to create squeezing and entanglement across two distinct degrees of freedom. We present our work in the specific physical context of a system consisting of collective atomic energy levels and discrete collective momentum states, but also consider other possible realizations. Our system uses a nonlinear Hamiltonian to generate dynamics in $SU(4)$, thereby creating the opportunity for dynamics not possible in typical $SU(2)$ one-axis twisting. This leads to three axes undergoing twisting due to the two degrees of freedom and their entanglement, with the resulting potential for a more rich context of quantum entanglement. The states prepared in this system are potentially more versatile for use in multi-parameter or auxiliary measurement schemes than those prepared by standard spin squeezing.
\end{abstract}

{
\let\clearpage\relax
\maketitle
}

\section{Introduction}

The creation and manipulation of quantum entanglement is central to developing powerful quantum technologies~\cite{QC_Preskill,AtomInt_Cronin}. In particular, precision measurements can greatly benefit from exploiting  quantum entanglements~\cite{LIGO_Tse,Gradiometry_Hardman} because non-classical states may be engineered for greater sensitivity to a parameter of interest compared to their classical counterparts~\cite{QMMet_Pezze,FermiGasMet_Luisa}. This field has seen rapid progress on several frontiers~\cite{AdvQMet_Giovannetti} including, but not limited to, experimental demonstration of atomic clock timekeeping below the shot noise limit~\cite{EntangledClock_Shu}, extensions of quantum error correction into quantum metrology schemes~\cite{QECMet_Lukin}, and machine learning and optimization for complex state preparation~\cite{RLMatter_Chih,QMVariation_Kaubruegger,OptimalMet_Marciniak} and measurement schemes~\cite{MLBayes_Pezze}. Through this rapid progress, there is the possibility that we will soon use quantum mechanical devices to probe new fundamental physics via tabletop experiments~\cite{RedShift_Bothwell,NewPhys_Safronova}.

Many state-of-the-art atom interferometry schemes rely on the process of spin squeezing~\cite{SqueezedAtom_Wineland,SpinSqueezeRydberg_Gil}, where a set of quantum spins are correlated to prepare a non-classical state that is sensitive to $SU(2)$ rotations at a precision below the standard quantum limit (SQL)~\cite{Squeezing_Ma} of $\Delta\phi^2 \propto 1/N$, where $\Delta\phi^2$ is the mean square error and $N$ is the number of particles used in the measurement. One candidate for generating this entanglement is one-axis twisting (OAT), whereupon many particles become entangled in a single degree of freedom under a non-linear Hamiltonian~\cite{Squeeze_Ueda,SqueezeNoise_Wineland}. Through entangling processes such as OAT, the SQL may be surpassed and a limit in precision of $\Delta\phi^2 \propto 1/N^2$ is achievable. This limit is a result of a Heisenberg uncertainty-like principle between the operator generating the unitary and the parameter one is measuring. This limit is aptly named Heisenberg limited scaling (HLS)~\cite{Bayes_Holland} and is the ultimate limit for metrological systems~\cite{QSensing_Degen}.

Schemes using OAT provide below SQL improvements for single parameter measurements, such as the angle a dipole sweeps under rotation generated by a magnetic field. These improvements are realized by sacrificing the variance of quantum fluctuations in one direction in exchange for a reduction in the variance of fluctuations in the direction we wish to measure. This hints at a natural extension of OAT; one where multiple degrees of freedom are entangled and squeezed to provide below SQL improvements for multiple parameters simultaneously.

In this paper, we introduce a novel method for squeezing and entangling two distinct degrees of freedom: the internal energy levels of an atomic ensemble and the collective atomic momentum. As a Gedanken experiment, we consider a collimated packet of atoms passing through a cavity. The cavity mediated emission and absorption of photons induces a twisting of the collective internal and momentum degrees of freedom, while also rapidly creating entanglement between these two degrees of freedom. The states prepared by this system could have the potential for multiparameter sensing and estimation~\cite{MultiParam_Pezze} below the SQL, squeezed state Bragg interferometry~\cite{Bragg_Shankar}, or single parameter estimation benefiting from  auxiliary measurements. By analyzing the Quantum Fisher Information Matrix (QFIM) of the system, we find that the maximum metrological gain in each individual degree of freedom is shown to scale proportionally to HLS. Here, we focus on the squeezing and correlation of the collective atomic internal energy state and momentum, but we emphasize that the general process could be realized with any system having same structure in its couplings and interactions. To this point, we discuss possible platforms which might be made to generate similar forms of entanglement in the conclusion of this paper.

The structure of this paper is as follows. In Section~\ref{sec:DerivHamil}, we cast the Hamiltonian into a form that illustrates the entanglement generating process: atomic emission and absorption of photons and the resulting momentum recoil. From this form, we show that some features may be intuitively understood as a generalization of the OAT Hamiltonian, while other important features have no analog in OAT. In Section~\ref{sec:OpAlg}, we explore the structure of the system and Hamiltonian using an underlying Lie algebra, and use these to simplify the subsequent analysis of the dynamics. In Section~\ref{sec:DynAnalysis}, we use the quantum Fisher information matrix (QFIM) to discuss the results of a numerical simulation of the time dynamics. Lastly, in Section~\ref{sec:InterScheme} we show schematically two interferometry protocols that benefit from the form of entanglement generated by this scheme.

\section{Derivation of the Hamiltonian and System Dynamics}\label{sec:DerivHamil}

\begin{figure}
\includegraphics[width=1\columnwidth]{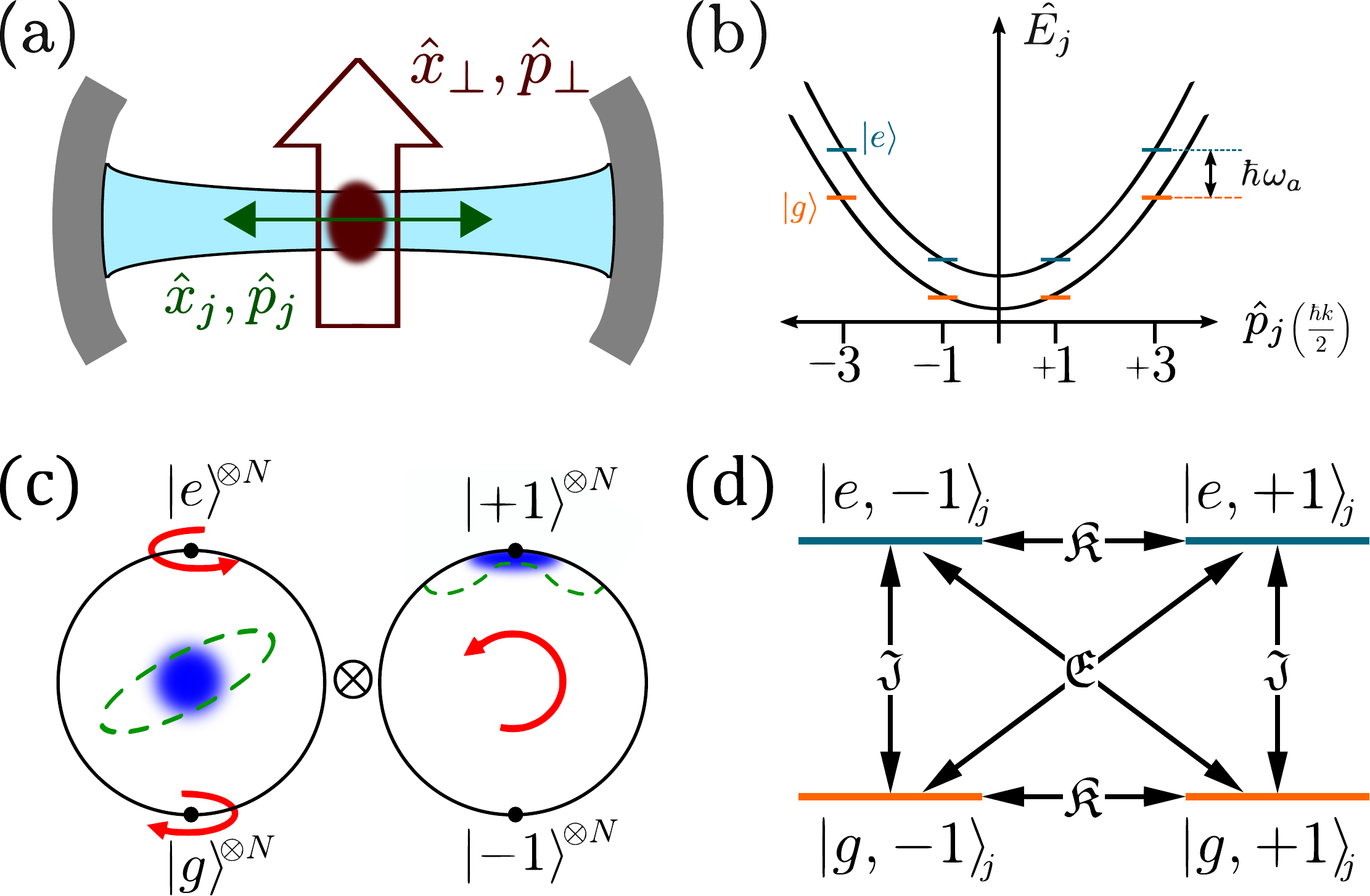}
\caption{\label{fig1} (a) Schematic of the proposed set up. Here, the momentum perpendicular to the cavity controls the interaction time. The initial momentum along the cavity axis selects the manifold of momentum states that the cavity couples to. (b) The spectrum of the kinetic energy versus the spectrum of momentum states. Here, we note the $\pm 3 \hbar k / 2$ states are far from the $\pm \hbar k / 2$ states, thus demonstrating that the lowest manifold of $4$ states can be considered isolated from the rest of the quadratic spectrum. (c) The two Bloch spheres for the collective two-level system. This picture is only valid when there is no entanglement between the two degrees of freedom, but it still provides a useful picture of the approximate behavior of the system. The blue cloud is the starting state, while the green dashed line represents the approximate distribution of the final state. The final state may not be fully represented on these Bloch spheres due to entanglement breaking the initial $SU(2)\otimes SU(2)$ symmetry needed to represent states on two collective Bloch spheres. (d) The four-level system, and black arrows representing each of the three unique $\ \mathfrak{su}(2)$ algebras acting on the system.}
\end{figure}

We consider the Gedanken experiment depicted in Fig.~\ref{fig1}(a), where a collimated packet of atoms passes through the center of the beam waist of a linear optical cavity, similar to a pulsed version of the set up proposed in~\cite{BeamLaser_Liu}. Each atom has a mass $m$, and two relevant internal energy levels labeled the excited and ground states $\ket{e}$ and $\ket{g}$, respectively. These energy levels are separated by the transition energy $\hbar\omega_a$. We assume that the cavity supports a single optical mode with corresponding frequency $\omega_{c}$, which is far detuned from the atomic transition by an amount $\Delta = \omega_a - \omega_c$. The interaction strength between the cavity photons and the $j$\textsuperscript{th} atom is taken to be $g(x_j)= \frac{g}{2} \cos(k \hat{x}_j)$. Furthermore, we assume $N$ atoms enter the cavity with uniform velocity, and spend a time $t$ inside the light-atom interaction volume. During this interaction time, the Hamiltonian is then: 

\begin{equation}
    \begin{aligned}
\hat{H}=&\sum_{j=1}^N \left(\frac{\hat{p}_j^2}{2m}+\frac{\hbar\omega_a}{2}\hat{\sigma}_j^z\right)+\hbar\omega_c\hat{a}_c^\dagger\hat{a}_c^{} \\
&+\frac{\hbar g}{2} \sum_{j=1}^N \cos(k\hat{x}_j) \left(\hat{a}_c^{}\hat{\sigma}^{+}_j+\hat{a}_c^{\dagger}\hat{\sigma}^{-}_j\right),
    \end{aligned}
\label{eq:Hfull}
\end{equation}
where $\hat{\sigma}^z_j=\ket{e}_j\bra{e}_j-\ket{g}_j\bra{g}_j$, $\hat{\sigma}^+_j=(\hat{\sigma}^-_j)^{\dagger}=\ket{e}_j\bra{g}_j$ are Pauli matrices for the  $j^{\text{th}}$ atom,  $\hat{p}_j$ ($\hat{x}_j$) is the transverse momentum (position) operator for the $j^{\text{th}}$ atom parallel to the cavity axis, and $\hat{a}_c^{\dagger}$ ($\hat{a}_c)$ is the photon creation (annihilation) operator of the cavity mode. 

The two relevant processes at play are the exchange of photons between different atoms and the atom's recoil due to the emission and absorption of photons. To simplify our study of these dynamics, we first take the interaction picture with $\hat{H}_0 =  \sum_{j=1}^N \hbar \omega_a \hat{\sigma}^z_j / 2 + \hbar \omega_a \hat{a}_c^\dagger \hat{a}_c $. We assume the cavity is in the dispersive regime $|\Delta| \gg \sqrt{N} g, \kappa$, where $\kappa$ is the cavity decay rate, such that we can adiabatically eliminate the cavity degrees of freedom over a coarse-grained timescale~\cite{LindbladEq_Jager}. The resultant Hamiltonian becomes
\begin{equation}
{\hat{H}}=\sum_{j=1}^N \frac{\hat{p}_j^2}{2m} + \sum_{i,j=1}^N \frac{\hbar g^2}{4 \Delta} \cos(k\hat{x}_i) \cos(k\hat{x}_j) \hat{\sigma}^{+}_i \hat{\sigma}^{-}_j.
\label{eq:Helim}
\end{equation}
The photon exchange has now been abstracted to an excitation exchange between different atoms and a resultant change in momentum. We note that the operators $\sum_{j=1}^N \cos(k\hat{x}_j) \hat{\sigma}^{\pm}_j $ cause a change in an atom's momentum by $\pm \hbar k $ upon trading an excitation, as $\exp( \pm i k \hat{x}_j )$ are the momentum shift operators. Therefore, if the atomic ensemble is prepared such that the atoms are in motional states differing in their momentum by integer multiples of $\hbar k$, the atoms will never leave this manifold under purely Hamiltonian evolution. We consider atoms in a superposition of motional states of the form $\ket{n}_j \equiv \ket{ n \hbar k / 2}_j$ for odd integers $n$. Preparation of such a state could be accomplished with a diffraction grating~\cite{atomOptics_Cronin} or via Kapitza-Dirac pulses and a trapping potential~\cite{Multimode_Smerzi}. 

Lastly, we assume that $ \hbar N g^2 / (4 \Delta) \ll (\hbar k)^2/m $, such that the lowest two momentum states are far detuned from the rest of the quadratic kinetic energy spectrum, as shown in Fig.~\ref{fig1}(b). Therefore, if the atoms start in the $\ket{\pm 1}_j$ states, they will in the subspace spanned by these two states. Under these conditions, the total kinetic energy remains fixed at $ N (\hbar k)^2 / (8 m) $. As a result, we can ignore the constant kinetic energy.

In this regime, the momentum now has a spin-$1/2$ algebraic structure and so the atom's momentum is effectively mapped onto a two-level system. We define $\hat{s}_{j}^{+}=(\hat{s}_{j}^{-})^{\dagger}=\ket{+1}_j\bra{-1}_j,$ and $\hat{s}_j^z = \ket{+1}_j\bra{+1}_j - \ket{-1}_j\bra{-1}_j $ such that we can cast the translation operator $\cos( k \hat{x}_j ) = [ \exp(ik\hat{x}_j) + \exp(-ik\hat{x}_j) ] / 2$ in terms of spin raising and lowering operators. We note that $ e^{+ik\hat{x}_j}=(e^{-ik\hat{x}_j})^{\dagger}=\hat{s}_{j}^{+}$ in this regime and therefore $2 \cos( k \hat{x}_j ) =  (\hat{s}_{j}^{+} + \hat{s}_{j}^{-}) \equiv \hat{s}^x_j $, thus we can rewrite our Hamiltonian in terms of these operators. Our simplified Hamiltonian therefore becomes
\begin{align} \label{eq:Hphys}
    \hat{H} &= \chi \sum_{i,j=1}^N\hat{s}_{i}^x\hat{s}_{j}^x\hat{\sigma}_{i}^+\hat{\sigma}_{j}^- ,
\end{align}
with $\chi = \hbar g^2/(16 \Delta)$. This non-linear Hamiltonian dictates how the atoms are to be entangled via cavity mediated interactions.

From Eq.~\ref{eq:Hphys}, we see that if the atoms enter the cavity in the same momentum state, with all atoms in the state $(\ket{+1}_j+\ket{-1}_j)/\sqrt{2}$, then the dynamics are generated by $\hat{H} \approx \sum_{i,j=1}^N\hat{\sigma}_{i}^+\hat{\sigma}_{j}^-\propto (\hat{J}^z)^2$, where $\hat{J}^z = \sum_j^N \hat{\sigma}_j^z/2$, and one-axis twisting is recovered. This is because the momentum flip operator, $\hat{s}^x_j$, affects an atom in the state $(\ket{+1}_j+\ket{-1}_j)/\sqrt{2}$ trivially. Physically, this is the case that all the atoms are in the same equal superposition of the $+\hbar k/2$ momentum states, so the recoil from emission and absorption of light doesn't affect the collective momentum, but the atom's internal degree of freedom remains free to evolve. With a starting state such as $\ket{+}^{\otimes N} =  (1/\sqrt{2})^N (\ket{e}+\ket{g})^{\otimes N}$ for the internal atomic energies, the Hamiltonian induces standard OAT behavior, leading to an effective spin squeezing. This starting state and behavior is shown in Fig.~\ref{fig1}(c), where the red arrows on the left Bloch sphere represent the action of $(\hat{J}^z)^2$.

We may also consider the case that the internal degrees of freedom don't affect the dynamics. This case is not physical, but rather provides an important intuition for the behavior in the system. Here, we take $\hat{H} \approx \chi \sum_{i,j=1}^N\hat{s}_{i}^x\hat{s}_{j}^x = 4 \chi (\hat{K}^x)^2$, where $\hat{K}^x = \sum_j^N \hat{s}_j^x/2$. While this is not necessarily physical, it sheds light on the approximate behavior of the atomic momentum: we expect the momentum states to experience OAT-like behavior through the non-linear rotation under $(\hat{K}^x)^2$. With a starting state of $\ket{+1}^{\otimes N}$ for the momentum degrees of freedom, we would expect to see operators orthogonal to $\hat{K}^x$, such as $\hat{K}^z = \sum \hat{s}^z_j / 2$, to undergo a twisting-like behavior. This starting state and approximate behavior is shown in Fig.~\ref{fig1}(c), where the red arrow on the right Bloch sphere represents the action of $(\hat{K}^x)^2$.

For the full Hamiltonian we expect the state $\ket{\psi}_0 = \ket{+}^{\otimes N} \otimes \ket{+1}^{\otimes N} $ to experience the corresponding spin twisting-like behavior in both degrees of freedom, and to lead to interesting entanglement between the two. In the subsequent sections, we demonstrate mathematically that this state breaks an important symmetry typically found in OAT, and then we numerically show this leads to entanglement that has potential for metrological advantage.

\section{The Operator Algebras}\label{sec:OpAlg}

In full, it is not immediately obvious how dynamics evolve under Eq.~\eqref{eq:Hphys}. The $\hat{s}^x_j$ operators complicate the Hamiltonian compared to the usual OAT Hamiltonian, preventing us from using methods typically used to solve OAT models. However, we can use the symmetries of the system to recast the Hamiltonian such that it is a member of an $\mathfrak{su}(4)$ algebra yielding a clear picture of the full dynamics and allowing for efficient numerical simulation.

The operators appearing in the Hamiltonian are all Pauli operators which correspond to a single atom's internal or momentum observable. For the $j^{th}$ atom's internal state, the operators $\{\hat{\sigma}^x_j,\hat{\sigma}^y_j,\hat{\sigma}^z_j\}$ fully describe any possible observable, where $\hat{\sigma}^x_j = \sigma^+_j + \sigma^-_j$ and $\hat{\sigma}^y_j = i( \sigma^-_j - \sigma^+_j)$. Similarly, its momentum state is fully described by $\{\hat{s}^x_j,\hat{s}^y_j,\hat{s}^z_j\}$, where $\hat{s}^y_j = i ( \hat{s}_{j}^{-} - \hat{s}_{j}^{+} )$ is needed for the momentum operators to close under commutation. The total system is then described, in part, by the collective atomic and momentum operators, $\hat{J}^i = \sum_j^N \hat{\sigma}_j^i/2$ and $\hat{K}^i = \sum_j^N \hat{s}_j^i/2$ for $i=x,y,z$ respectively. These collective atomic and momentum operators each form an $\mathfrak{su}(2)$ algebra: $\mathfrak{J}=\{\hat{J}^z,\hat{J}^\pm\}$ and $\mathfrak{K}=\{\hat{K}^z,\hat{K}^\pm\}$. These two algebras allow us to fully describe any state which is seperable in the two degrees of freedom, such as the state $\ket{\psi_0}$ which is represented on two composite Bloch spheres in Fig.~\ref{fig1}(c) in blue. Importantly, we note that the momentum operator $\hat{K}^z$ corresponds to the observable for the center of mass momentum, $\hat{P}_{\rm COM} = \hbar k \hat{K}^z$, which is intuitively the difference between the number of atoms moving in the $+1$ and $-1$ eigenstates.

We can further simplify our analysis by mapping particles into the Schwinger boson representation~\cite{Schwinger}. Here we use the simultaneous eigenstates of $\hat{J}^z$ and $\hat{K}^z$ as the basis for the new representation, but in general this could be done via the procedure shown in~\cite{SUNBos_Manu}.
First, we define
\begin{equation}
    \begin{aligned}
 \label{eq:SchwingState}
    & \ket{\alpha,\beta,\gamma,\delta} = \\
    & \mathcal{S}\left(\ket{e,+1}^{\otimes \alpha} \ket{g,-1}^{\otimes \beta} \ket{e,-1}^{\otimes \gamma} \ket{g,+1}^{\otimes \delta}\right), 
    \end{aligned}
\end{equation}
where $\alpha+\beta+\gamma+\delta=N$ is the total number of atoms and $\mathcal{S}$ is the symmeterization operator. Note that the symmetrizer is defined with the normalization factor, shown explicitly in Appendix~\ref{sec:Normalization}, so this representation is normalized. We can represent all the relevant operators in this formalism as well by associating the annihilation (creation) operators $\hat{a},\hat{b},\hat{c},\hat{d}$ ($\hat{a}^{\dagger},\hat{b}^{\dagger},\hat{c}^{\dagger},\hat{d}^{\dagger}$) to each of the four modes, such that $\hat{a} \ket{\alpha,\beta,\gamma,\delta} = \sqrt{\alpha} \ket{\alpha-1,\beta,\gamma,\delta} $
and similarly for the other three modes as shown in Appendix~\ref{sec:create_annihilate}. Now, the number of atoms in the excited state is simply $\alpha + \gamma$ for states of the form in Eq.~\eqref{eq:SchwingState}. Therefore, we define $\hat{n}_e \ket{\alpha,\beta,\gamma,\delta}  = ( \hat{a}^{\dagger}\hat{a} + \hat{c}^{\dagger}\hat{c} ) \ket{\alpha,\beta,\gamma,\delta}$. By the same process, we can recover the ground state number operator to be $\hat{n}_g = \hat{b}^{\dagger}\hat{b} + \hat{d}^{\dagger}\hat{d}$, the $+1$ momentum state number operator to be $\hat{n}_{+1} = \hat{a}^{\dagger}\hat{a} + \hat{d}^{\dagger}\hat{d}$, and the $-1$ momentum state number operator to be $\hat{n}_{-1} = \hat{b}^{\dagger}\hat{b} + \hat{c}^{\dagger}\hat{c}$. Our collective atomic and momentum operators are simple to represent in the form
\begin{equation} 
    \begin{aligned}
\hat{J}^z & = \frac{1}{2} ( \hat{n}_e - \hat{n}_g ), \\
\hat{K}^z & = \frac{1}{2} ( \hat{n}_{+1} - \hat{n}_{-1} ),
    \end{aligned}
\end{equation}
and $\hat{J}^{-} =\hat{a}\hat{d}^\dagger + \hat{c}\hat{b}^\dagger = ( \hat{J}^{+} )^{\dagger}, 
\hat{K}^{-} =\hat{a}\hat{c}^\dagger + \hat{d}\hat{b}^\dagger = ( \hat{K}^{+} )^{\dagger}$. Moreover, the Hamiltonian is also simply represented,
\begin{equation} 
\hat{H} = \chi ( \hat{a}^{\dagger} \hat{b} + \hat{c}^{\dagger} \hat{d} ) ( \hat{a} \hat{b}^{\dagger} + \hat{c} \hat{d}^{\dagger} ).
\end{equation}
This is intuitively what should be expected because, for example, $\hat{a} \hat{b}^{\dagger}$ is collective emission where a single atom goes from the excited, +1 motional state to a ground, -1 motional state. The other terms can be similarly understood.

Lastly, we introduce the raising and lowering operators $\hat{E}^+ = \hat{a}^{\dagger} \hat{b} + \hat{c}^{\dagger} \hat{d} = (\hat{E}^-)^\dagger $, and we notice that $ [ \hat{E}^+, \hat{E}^- ] = 2 \hat{J}^z $ and 
$[ \hat{J}^z, \hat{E}^\pm ] = \pm \hat{E}^\pm .$ Thus, we see that the set $\mathfrak{E} = \{\hat{J}^z,\hat{E}^\pm\}$ forms a third closed $\mathfrak{su}(2)$ algebra on the system which succinctly represents the entanglement generating processes due to absorption and emission. The three sub-algebras $\mathfrak{J},\mathfrak{K}$ and $\mathfrak{E}$ taken together are members of a complete $\mathfrak{su}(4)$ algebra, which generates an $SU(4)$ group that efficiently describes the dynamics of this system. The action of three sub-algebras is represented schematically in Fig.~\ref{fig1}(d) for a single atom. In summary, within the full $\mathfrak{su}(4)$ describing our dynamics, we find that there exists three $SU(2)$ subgroups each generated by $\mathfrak{J},\mathfrak{K}$, or $\mathfrak{E}$, which matches the general structure for $SU(4)$~\cite{SUN_Yukawa}. Thus, the system can be considered as a collection of hybridized angular momentum.

We can take advantage of the commutation structure in $\mathfrak{E}$ to simplify the Hamiltonian even further,
\begin{equation}        \begin{aligned} \label{eq:HSU4}
\hat{H} &= \chi \hat{E}^+ \hat{E}^- \\
&= \chi ( \hat{E}^2 - (\hat{J}^z)^2 + \hat{J}^z ),
    \end{aligned}
\end{equation} 
where $ \hat{E}^2 = \hat{E}^+ \hat{E}^- + (\hat{J}^z)^2 - \hat{J}^z $ is the quadratic Casimir operator~\cite{RepThrySU2_Rocek} for $\mathfrak{E}$. Now, Eq.~\eqref{eq:HSU4} looks like the familiar form of a OAT Hamiltonian, except for the important difference that $\hat{K}^y$, $\hat{K}^z$ don't commute with $\hat{E}^2$. This means there exists states which are eigenstates of $\hat{K}^z$ that evolve non-trivially under the operator $\hat{E}^2$, such as the starting state discussed at the end of Section~\ref{sec:DerivHamil}. Furthermore, we can observe that the operator $\hat{E}^2$ has shells corresponding to each of its eigenvalues, similar to the shells typically defining eigenvalues for total angular momentum observables. The starting state, $\ket{\psi_0}$ creates a superposition over these shells and, with $\hat{E}^2$ contributing non-trivially to the dynamics, each of the three pseudo-angular momentum subgroups experience a twisting under this Hamiltonian.

\section{Analysis of the Dynamics and Entanglement Generation}\label{sec:DynAnalysis}
Now we use the Schwinger boson representation introduced in Section~\ref{sec:OpAlg} to numerically simulate the system and explore the dynamics. For these simulations we assume the cavity decay at rate $\kappa$ and other dissipative processes, such as spontaneous emission at rate $\gamma$, are negligible. This assumption is valid in the limit that the timescale considered for unitary dynamics, $t$, is much smaller than the relevant inverse decay rates. Further analysis of the effects of decoherence is left to future work, but we attempt to make explicit note of when one would expect decoherence to become non-negligible, and the relevant bounds in these cases.

To simulate the system, we use the four annihilation/creation operators found in the previous section, and model the atomic system as a system of four harmonic oscillators. The Hilbert space of four harmonic oscillators has a dimensionality of $(N+1)^4$ containing all states with atom numbers between $0$ and $4N$ atoms. We may use either of the conditions that $\hat{n}_e+\hat{n}_g = N$ or $\hat{n}_{+1}+\hat{n}_{-1} = N$ to project onto only the states with $N$ atoms. This corresponds to restricting to only states which are may be reached by $SU(4)$ action, and the typical argument of putting $N$ atoms indistinguishably in four distinguishable states shows the system scales at $(N+1)(N+2)(N+3)/6$ states for $N$ atoms. This now matches the dimensionality of the basis states with an $SU(4)$ symmetry, given in Ref~\cite{SU4_Xu}, and is numerically more efficient than the initial $(N+1)^4$ scaling.

We use the starting state discussed in Section~\ref{sec:DerivHamil}, $\ket{\psi}_0 = \ket{+}^{\otimes N} \otimes \ket{+1}^{\otimes N}$. As noted in the end of Section~\ref{sec:OpAlg}, $\ket{\psi}_0$ is not an eigenstate of $\hat{E}^2$. From the discussion of this state and the picture in Fig.~\ref{fig1}(c), we expect this initial state to lead to twisting-like behavior and entanglement generation between the two degrees of freedom. The intuitive picture to understand this behavior is the following. When an atom emits light, its internal degree of freedom becomes entangled to that of the atom which absorbs the emitted light. At the same time, both these atom's momentum states must switch, causing their external degrees of freedom to become entangled similar to their internal ones.

To diagnose the amount of entanglement and potential metrological use, we consider the case that one wants to prepare states which will be used to estimate some phase, $\phi_j$, encoded by unitary evolution under some operator, $\hat{G}^j$, so that the state evolves according to $\exp(-i \phi_j \hat{G}^j )$. Specifically we consider the cases that $\hat{G}^j$ is in either $\mathfrak{J}$ or $\mathfrak{K}$, and choose the indices $i,j$ so that if $i,j=1,2,3$ then $\hat{G}^i, \hat{G}^j = \hat{J}^x, \hat{J}^y,\hat{J}^z$ and if $i,j = 4,5,6$ then $\hat{G}^i, \hat{G}^j = \hat{K}^x, \hat{K}^y, \hat{K}^z$. In this scenario the QFIM serves as both an entanglement measure~\cite{Entangle_Pezze} and a measure of the potential metrological use of a state in quantum metrology~\cite{QFIMrev_Liu}. We use for the form of the QFIM given in Ref.~\cite{QFIM_Meyer} for pure states, since in the present proof of concept we do not address decoherence. Under this condition, the matrix elements are given by
\begin{eqnarray} \label{eq:QFIM}
\mathcal{F}^{ij} = 4 \Big( \Big\langle \frac{\{\hat{G}^i,\hat{G}^j\}}{2} \Big\rangle - \langle \hat{G}^i \rangle \langle \hat{G}^j \rangle \Big),
\end{eqnarray}
where $\{\hat{G}^i,\hat{G}^j\} = \hat{G}^i \hat{G}^j + \hat{G}^j \hat{G}^i $ is the anti-commutator. For $i=j$, Eq.~\eqref{eq:QFIM} returns the fourfold variance, which captures the amount of squeezing and entanglement present. The condition for an entanglement witness to squeezing is $\mathcal{F}^{ii} / N^2 > 1/N$, which is equivalent to the condition given in Ref.~\cite{Entangle_Pezze}. If $\mathcal{F}^{ii} / N^2$ approaches a constant as $N$ grows, then the sufficient condition for entanglement is clearly met and the system's potential metrological use proportional to HLS is demonstrated. Meanwhile for $i\neq j$, Eq.~\eqref{eq:QFIM} returns the fourfold covariance, thereby capturing the amount of quantum correlations between these observables. We observe that $[\hat{J}^i, \hat{K}^j] = 0$ for all $\hat{J}^i\in\mathfrak{J}, \hat{K}^j\in\mathfrak{K}$. As a result the covariance of two operators on the internal state and the momentum, such as $\text{cov}(\hat{J}^x, \hat{K}^z)$, is non-zero only for pure states which are entangled. The off-diagonal elements of the QFIM with $i \in \{ 1,2,3 \}$ and $j \in \{ 4,5,6 \}$ therefore represents the covariance between the atomic and momentum operators, and acts as an entanglement witness of quantum correlations between the two degrees of freedom. Thus, we use the sufficient condition that $\mathcal{F}^{ij} \neq 0$ as an entanglement witness for the two degrees of freedom as a pure state bipartite system. This is a modified version of the condition given in Ref.~\cite{Covar_Abascal}.

\begin{figure}
\includegraphics[width=1\columnwidth]{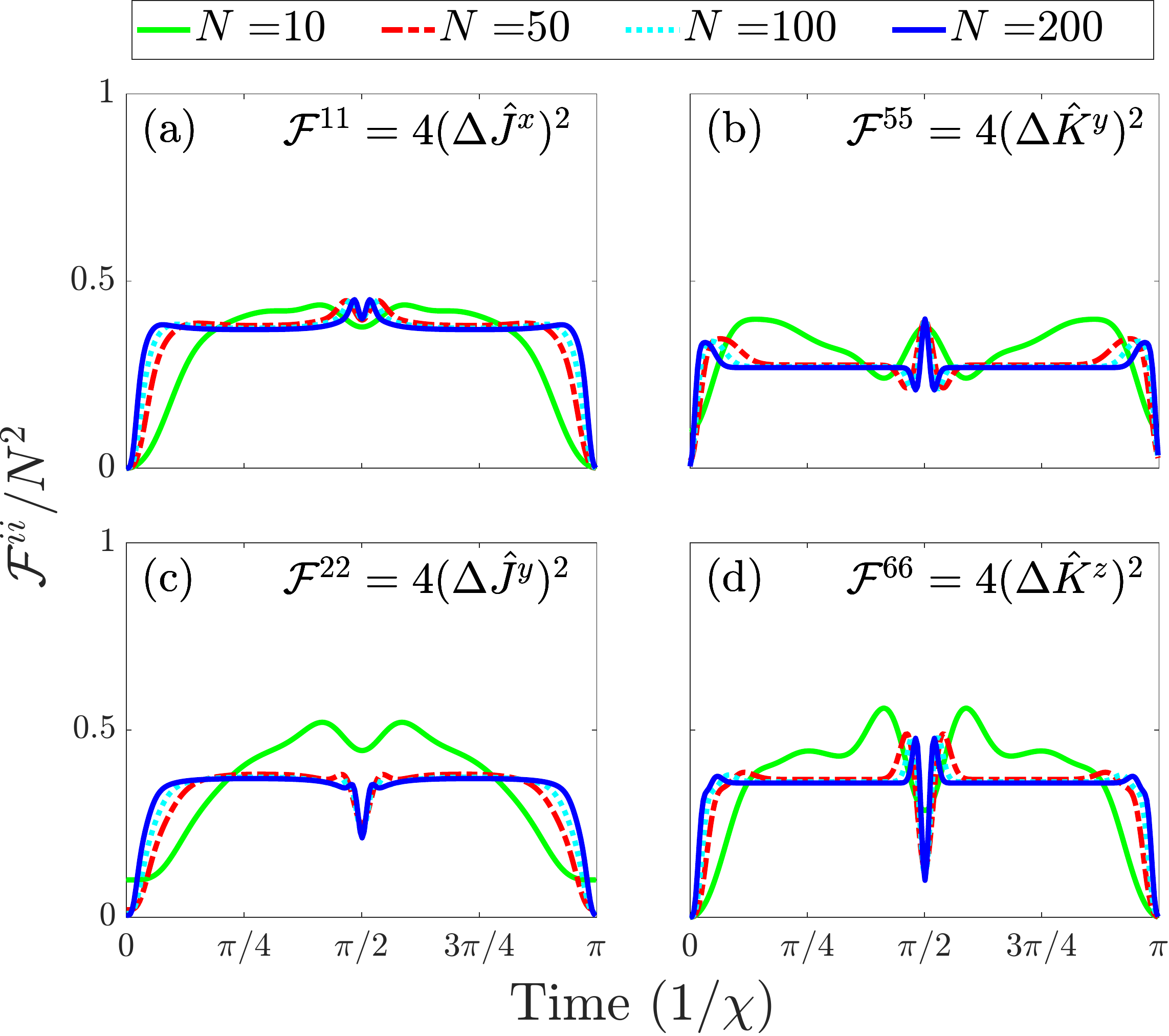}
\caption{\label{fig2} Four of the six diagonal elements of the QFIM, for four different atom numbers. The operators $\hat{J}^z$ and $\hat{K}^x$ are left out because they commute with the Hamiltonian and are therefore conserved quantities. We see that as the number of atoms grow, the behavior of the diagonal QFIM elements converge. For atom numbers of $N\approx50$ or more, a plateau with respect to time appears, centered around $\chi t = \pi / 4$. This is similar to the behavior found in OAT where the QFIM for $\hat{J}^x$ and $\hat{J}^y$ reach a plateau~\cite{QMMet_Pezze} centered around the same time. As $N$ grows, the plateau exists almost everywhere in time. Here we only show even atom numbers, $N$, but we note that for odd atom numbers the behavior is the same except for at $\chi t = \pi/2$, where the concavity is opposite from what's shown here.}
\end{figure}

In Fig.~\ref{fig2}, we show the quantity $\mathcal{F}^{ii}/N^2$ for the four operators of interest, and for four different numbers of atoms, $N$, each as a function of interaction time with the cavity, $t$. We observe that $\mathcal{F}^{ii} / N^2$ increases sharply before leveling off to a constant value over time. Because $\mathcal{F}^{ii} / N^2 > 1/N$, the entanglement witness condition is satisfied for each case. This condition is met in a short interaction time, demonstrating that entanglement is quickly generated in both the collective internal and momentum modes. Therefore we see that along with the spin squeezing in the internal atomic degrees of freedom, this platform also leads to an effective squeezing of the momentum degrees of freedom.

\begin{figure}
\includegraphics[width=1\columnwidth]{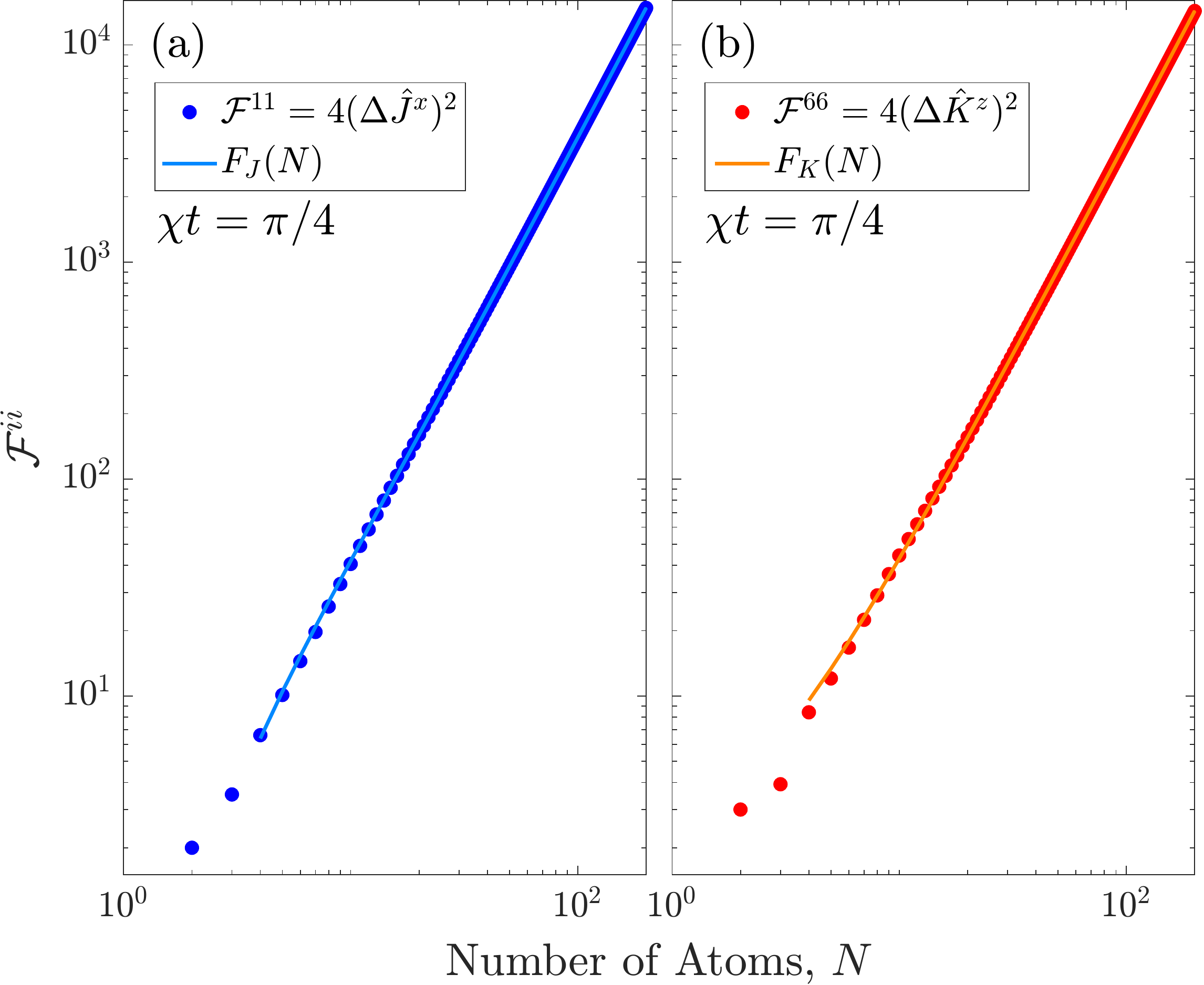}
\caption{\label{fig3} The diagonal elements of the QFIM corresponding to $\hat{J}^x$ and $\hat{K}^z$ shown as a function of atom number, $N$. We fit $4 \Delta\hat{J}^x$ and $4 \Delta\hat{K}^z$ with second order polynomials $F_J(N)$ and $F_K(N)$ respectively. We fit for $N\geq4$, because for $N=2$ and $N=3$ the system has anomalous behavior for small atom numbers. We find that $4 \Delta\hat{J}^x$ is fit with the function $F_J(N) \approx 0.366 N^2 +  0.793 N - 2.662 $, and $4 \Delta\hat{K}^z$ is fit with the function $F_K(N) \approx 0.356 N^2 +  0.599 N + 1.466$. Both of these demonstrate the HLS.}
\end{figure}

To quantify the potential metrological use of this system, we fix the time at $\chi t = \pi / 4$ and show how the diagonal element of QFIM for $\hat{J}^x$ and $\hat{K}^z$ scales with atom number. The results are shown in Fig.~\ref{fig3}. Achieving an interaction time scale of $\chi t \sim 1$ would require a very big cavity-atom coupling, such that $\chi \gg \gamma$. The same is true for any other decoherence rate one might consider. As a result, this timescale may be physically inaccessible with traditional cavities, but is theoretically interesting nonetheless. These long timescales form the equivalent of the ``oversqueezed" timescales in standard OAT. We specifically choose the time $\chi t = \pi / 4$ because it is the center of the plateau in the QFIM's diagonal elements.

We see that both the atomic and momentum degrees of freedom scale proportionally to $N^2$, i.e. with HLS. Similar behavior exists in OAT, where one finds a plateau in the variance of the antisqueezed quadrature for times between $1/\sqrt{N} \lesssim \chi' t \lesssim \pi/2 - 1/\sqrt{N}$, where $\chi'$ is an appropriately defined frequency. However, in OAT this plateau restricted to just the spin degree of freedom~\cite{QMMet_Pezze}. Our scheme provides a squeezing mechanism for momentum degrees of freedom, creating the possibility that spin-squeezing techniques used in Ramsey interferometry \cite{Clock_Lukin} might be generalized to Bragg interferometry or that the two might be performed simultaneously.
 
\begin{figure}
\includegraphics[width=1\columnwidth]{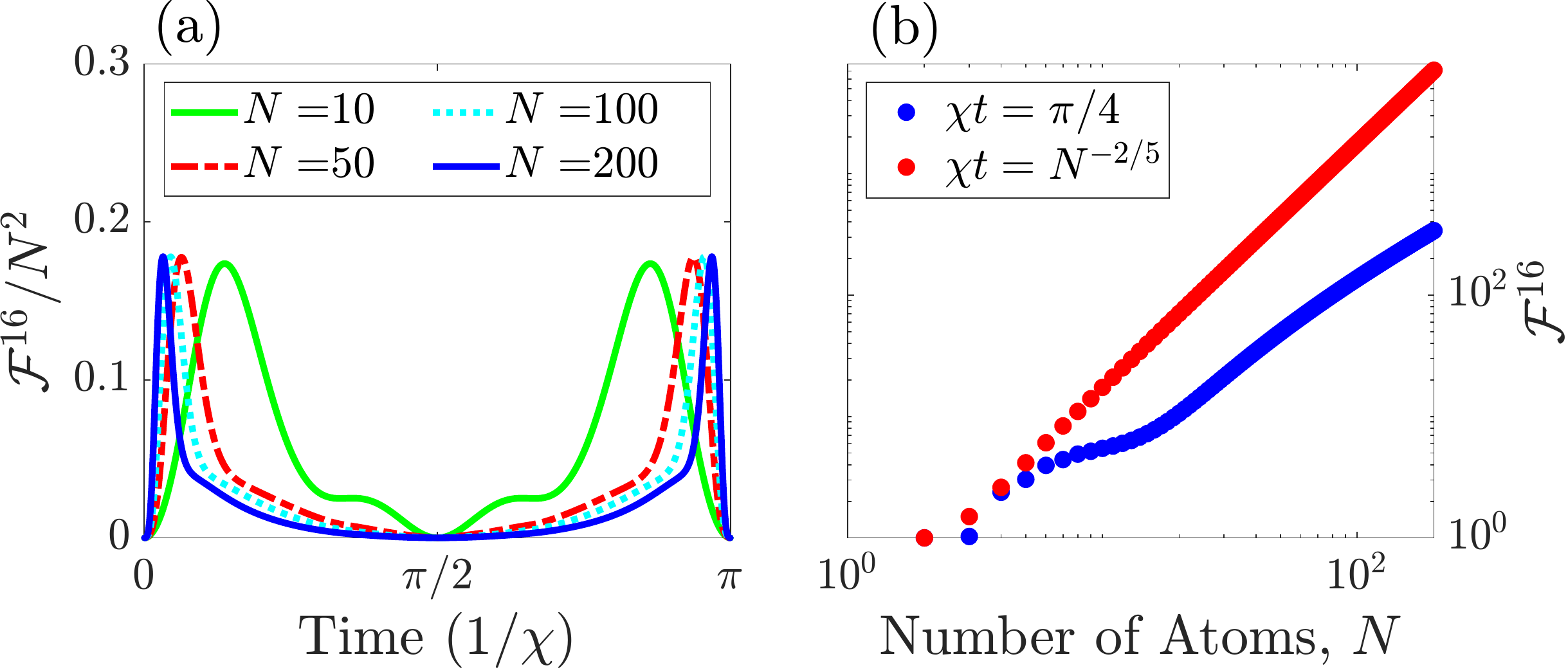}
\caption{\label{fig4} Plots of $\mathcal{F}^{ij} = 4 \mathrm{cov}(\hat{J}^x,\hat{K}^z) $. Left - The off diagonal of the QFIM, $\mathcal{F}^{ij}$, normalized by $N^2$ for four different values of $N$. We see the covariance between $\hat{J}^x,\hat{K}^z$ grows rapidly before decaying for longer time scales, then in a collapse-revival like effect at $\chi t \approx \pi$ the operators become correlated again before approaching the starting state. Right - The same off diagonal element of the QFIM at two different times: $\chi t = \pi/4$ when the correlations are decreasing, and $\chi t = N^{-2/5}$ when the correlations are largest. We find that $\mathcal{F}^{ij}|_{\chi t = \pi / 4} \approx 4.103\cdot10^{-3} N^2 + 0.926 N $, and $\mathcal{F}^{ij}|_{\chi t = N^{-2/5}} \approx 0.1782 N^2 - 0.02721 N $ }.
\end{figure}

Now, we study the behavior of the entanglement between the degrees of freedom, which has no analog in OAT. We study the entanglement via the fourfold covariance between the two operators $\hat{J}^x$ and $\hat{K}^z$, corresponding to an off-diagonal element of the QFIM. In Fig.~\ref{fig4}(a), we see that the system moves through a highly correlated state, with a high covariance between the two degrees of freedom, before it approaches an uncorrelated state for a moment in time at $\chi t = \pi/2$. At an interaction time of $\chi t = \pi$, the system returns to its original state. In Fig.~\ref{fig4}(b), we see that for interaction times of $\chi t \approx \pi / 4$ the correlations only scale linearly with $N$. Therefore, interaction times which reach this plateau prepare a system which is capable of quantum sensing for two parameters at the Heisenberg limit, with relatively little error introduced by the simultaneous measurement of the two parameters. This motivates the first half of the next section, where a schematic representation of two parameter interferometry is shown.

The time at which the system is maximally correlated is labeled $t_{\text{max}}$, and we find $\chi t_{\text{max}}$ decreases with number of atoms such that $\chi t_{\text{max}} \approx N^{\nu}$, where $\nu \approx -2/5$ is found from fitting. At this time, the maximum correlation scales proportionally to $N^2$, which is on the order of the squeezing for the two degrees of freedom. 

To achieve an interaction time with these large correlations, one needs $\chi t\sim N^{-2/5}$. Compared to the single particle emission, one has the requirement $\chi \gg N^{-2/5}\gamma$, which can be achieved for sufficiently large $N$. Therefore we expect single-particle decoherence to be negligible on these timescales. In this regime, we instead expect that collective decoherence processes, such as collective spontaneous emission mediated by the cavity, would limit the amount of achievable entanglement. After adiabatic elimination of the cavity, the collective decoherence rate is due to light being incoherently scattered into the cavity and lost. This rate can be estimated as $ N \chi \kappa / \Delta \propto N g^2 \kappa / \Delta^2 $. Therefore one may reduce it by increasing the cavity-atom detuning, $\Delta$, at the expense of reducing $\chi$. However, interaction times of $\chi t\sim N^{-2/5}$ may still be possible in cavities with low photon loss rate $\kappa$. We also note that a similar timescale of $\chi t\sim N^{-2/3}$ is needed for production of optimally squeezed states in standard OAT~\cite{QMMet_Pezze,RobustSpinSqueeze_Rey}, so it could be possible to achieve an interaction time on the order needed to see strong correlations.

This short timescale with highly correlated degrees of freedom motivates the second half of our next section, where a schematic representation of single parameter interferometry is shown. The parameter is estimated via an interaction with one degree of freedom, and an auxiliary measurement on the other degree of freedom.

\section{Interferometer Schemes}\label{sec:InterScheme}

To demonstrate a possible metrological use, we numerically explore two interferometry schemes. The first uses the system to detect two relative phases: one encoded in the atom's internal degree of freedom, and a second encoded in the momentum degree of freedom. The second scheme uses this system to detect a single parameter via auxiliary measurements. The version of the auxiliary measurement scheme presented here is the case that the collective internal degree of freedom accumulates phase and the momentum degree of freedom is measured. However, this process would work similarly if the roles were reversed.

For both schemes, we choose a new interaction picture for the Hamiltonian such that $\hat{J}^z$ is removed from Eq.~\eqref{eq:HSU4}. This has no effect on the physics described above, besides keeping the atomic ensemble polarized in $\hat{J}^x$ instead of precessing about $\hat{J}^z$. This matches what is often done in OAT, and the process is shown in more depth in Appendix~\ref{sec:InterHamil}.

\begin{figure}[h]
\mbox{ 
\Qcircuit @C=1em @R=0.7em {
\lstick{(\text{a})\ \ } & & & & \\
& \lstick{\ket{+}^{\otimes N}} & \multigate{1}{e^{-i \hat{H} \tau_2}} & \gate{e^{-i \theta_{\text{opt}} \hat{J}^x}} & \gate{\ e^{-i \phi_3 \hat{J}^z} \ } & \measureD{\hat{J}^x} \\
& \lstick{\ket{+1}^{\otimes N}} & \ghost{e^{-i \hat{H} \tau_2}} & \qw & \gate{\ e^{-i \phi_5 \hat{K}^y} \ } & \measureD{\hat{K}^z} \\ \\ \\
\lstick{(\text{b})\ \ } & & & & \\
& \lstick{\ket{+}^{\otimes N}} & \multigate{1}{e^{-i \hat{H} \tau_1}} & \gate{e^{-i \phi_1 \hat{J}^x}} & \multigate{1}{e^{i \hat{H} \tau_1}} & \qw \\
& \lstick{\ket{+1}^{\otimes N}} & \ghost{e^{-i \hat{H} \tau_1}} & \qw & \ghost{e^{i \hat{H} \tau_1}} & \measureD{\hat{K}^z}
} }
\caption{\label{fig5} A quantum circuit schematic of the two schemes. The two tracks represent the actions affecting either degree of freedom, with the top track representing the internal states of the atoms, and the bottom track representing the momentum. (a) The two parameter scheme. The interaction time for this two parameter scheme, $\tau_2$, is fixed at $\chi \tau_2 = \pi / 4$ to demonstrate metrological use on the platuea found in Section~\ref{sec:DynAnalysis}. (b) The auxiliary measurement scheme. Here, $\chi \tau_1 = N^{-2/5}$ is chosen such that the ensembles are maximally correlated. The time-reversed unitary could be achieved by changing the detuning on the cavity.}
\end{figure}
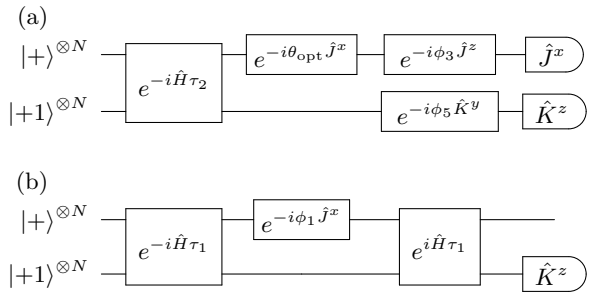

\begin{figure}
\includegraphics[width=1\columnwidth]{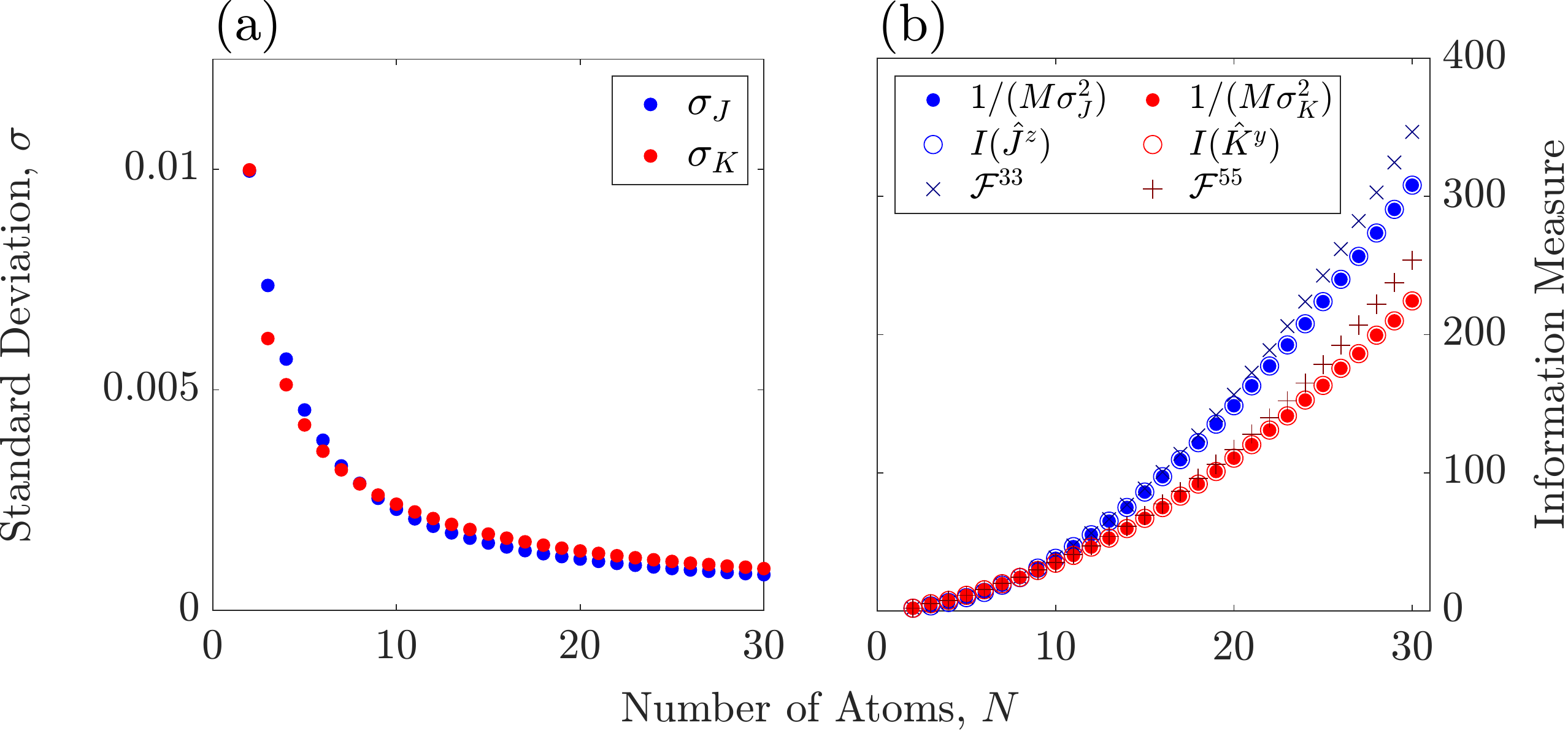}
\caption{\label{fig6} Left - A plot of the standard deviation corresponding to the final result of Bayesian inferencing for estimating the phases $\phi_3$ and $\phi_5$ with $M = 5000$ measurements, and $\phi_3=\phi_5=\pi/16$. Right - A plot of the quantities $\frac{1}{M \sigma_i^2}$ for $\sigma_i = \sigma_J, \sigma_K$, the CFI $I(\hat{G}^i)$ for $\hat{G}^i = \hat{J}^z, \hat{K}^y$ corresponding to these measurements, and the diagonal elements of the QFIM for these measurements. Note that because of the rotation generated by $\hat{J}^x$ prior to the interferometry, $\mathcal{F}^{33} = (\Delta \hat{J}^z)^2$ now scales with HLS. We see that the quantities $\frac{1}{M \sigma_i^2} = I(\hat{G}^i)$ saturate the classical CRB from the left half of Eq.~\eqref{eq:CRB_inv}, and nearly saturate the quantum CRB. By fitting the diagonal QFIM elements and $\frac{1}{M \sigma_i^2}$ we find the CFI scales as $I(\hat{J}^z) \approx 0.3184 N^2 + 0.9162  N $, $I(\hat{K}^y) \approx 0.2022 N^2 + 1.454  N $, while $\mathcal{F}^{33} = (\Delta \hat{J}^z)^2 \approx 0.3815 N^2 + 0.1577 N$, $\mathcal{F}^{55} = (\Delta \hat{K}^y)^2 \approx 0.2512 N^2 + 0.8727 N$. This indicates this measurement scheme scales at about $80\%$ the theoretical maximum.}
\end{figure}

We start with the two parameter scheme. The relevant schematic representation is shown in Fig.~\ref{fig5}(a). Here, we first pass the atomic ensemble through the cavity for an interaction time $\chi \tau_2 = \pi / 4$ to prepare the probe state. We chose this time to show the metrological use for times near the plateau, when correlations between the degrees of freedom are decreasing with respect to interaction time. However, this multiparameter scheme could be used for any interaction time, albeit with slight differences due to varying correlation strengths. After the state preparation, a rotation generated by $\hat{J}^x$ is performed so that the maximum fluctuation is in $\hat{J}^z$, where the angle $\theta_\text{opt}$ is found numerically. For the momentum degree of freedom, it was found that the state is already prepared such that the maximal fluctuations are along $\hat{K}^y$ at this time. The signal is encoded in the system by unitary
\begin{equation} \label{eq:Signal}
    \hat{V} = \exp( - i \phi_3 \hat{J}^z - i \phi_5 \hat{K}^y ),
\end{equation}
where we assume for numerical convenience the phases $\phi_3,\phi_5$ are small, at $\phi_3=\phi_5=\pi/16$. However, we found that these results hold for larger phases as well as for two phases which aren't equal. After the unitary, we measure the observables $\hat{J}^x$ and $\hat{K}^z$ and carry out phase estimation for both phases simultaneously. To estimate the phase, we simulate a Bayesian inferencing scheme~\cite{Bayes_Holland} for two parameters and with a flat prior, and to find the asymptotic behavior of this Bayesian inference, we numerically calculate the Classical Fisher Information (CFI) as a function of atom number. The exact algorithm for sampling and updating a probability distribution, as well as the explicit form of the CFI are shown in Appendix~\ref{sec:Bayes}. Using the CFI, we have a useful set of inequalities from the Cram\'er-Rao Bound~\cite{StatDist_Caves} (CRB):
\begin{equation} \label{eq:CRB_inv}
    \sigma_i^2 \geq \frac{1}{M I(\hat{G}^i)} \geq \frac{1}{M \mathcal{F}^{ii}}
\end{equation}
where $i=3,5$ corresponds to either $\phi_3,$ or $\phi_5$, $\sigma_i^2$ is the variance of the probability distribution, $M$ is the number of measurements, $I(\hat{G}^i)$ is the CFI for a parameter encoded by the operator $\hat{G}^i = \hat{J}^z,\hat{K}^y$, and $\mathcal{F}^{ii}$ is the diagonal element of the QFIM for the corresponding operator. The first inequality is the classical CRB, and the second inequality is the quantum CRB. By inverting this bound we find the following: $ \mathcal{F}^{ii} \geq I(\hat{G}^i) \geq \frac{1}{M \sigma_i^2}$, so we can tell how close our resultant probability distribution from Bayesian inferencing is to saturating the CRB. In Fig.~\ref{fig6}, we see the results of this analysis for $M=5000$ measurements. This measurement scheme saturates the classical CRB for both parameters, and reaches a value of about $80\%$ of the quantum CRB. Moreover, it does this simultaneously for both parameters. 

We also note that, while not shown, as $\phi_3,\phi_5$ tend towards zero the CFI exactly saturates the quantum CRB, but Bayesian inferencing takes asymptotically more measurements to saturate the classical CRB. This result was found numerically, but it can be intuitively explained by the formation of narrow, ring-like $Q$ functions on the collective Bloch spheres of the internal and external degrees of freedom. Those rings form along the $J^x$-$J^z$ plane and along the $K^y$-$K^z$ plane which makes them sensitive to any rotation which results in leaving the corresponding planes. For rotations of these planes around the $J^z$ and $K^y$ axes one can then efficiently read out the applied phase by measuring $J^x$ and $K^z$, respectively. With this picture in mind, we would expect the optimal measurement for any value of $(\phi_3,\phi_5)$ is $(\hat{J}^x \cos(\phi_3) + \hat{J}^y \sin(\phi_3) ) \otimes (\hat{K}^z \cos(\phi_5) + \hat{K}^x \sin(\phi_5) )$, such that the measurement will always be oriented the same relative to plane this state is in. Numerically we find that this is in fact always saturates the quantum CRB. However, using this measurement requires knowledge of $(\phi_3,\phi_5)$.

\begin{figure}
\includegraphics[width=1\columnwidth]{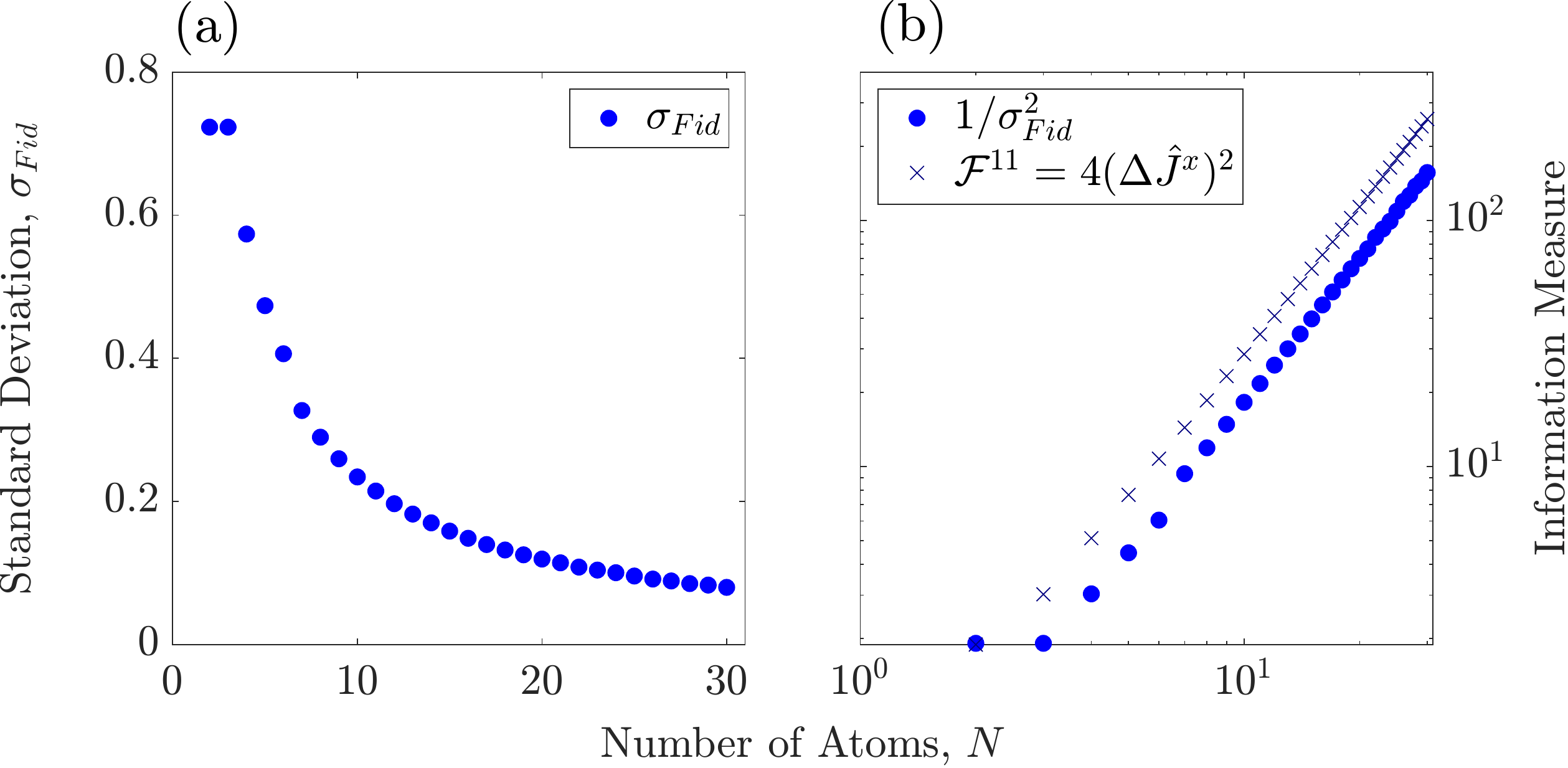}
\caption{\label{fig7} Left - The standard deviation of the final state fidelity, $\sigma_{Fid}$ with the $\ket{+1}^{\otimes N}$ momentum state. This is found by fitting the central peak with a Gaussian and offset. Right - The quantities $1/\sigma_{Fid}^2$ and the QFIM element corresponding to rotations about $\hat{J}^x$. We see that $1/\sigma_{Fid}^2 \approx 0.1699 N^2 + 0.1069 N$ and $\mathcal{F}^{ii}|_{\chi t = N^{-2/5}} \approx 0.2874 N^2 - 0.0577 N$, showing that this auxiliary measurement reaches about $0.6$ the quantum CRB. }
\end{figure}

Now, we turn our attention to the auxiliary measurement scheme, shown in Fig~\ref{fig5}(b). Here, the atomic ensemble first passes through the cavity for a time of $\chi \tau_1 = N^{-2/5}$, so that the observables $\hat{J}^x$ and $\hat{K}^z$ are well correlated. Then, the phase is encoded on either the internal degree of freedom or the momentum. By changing the detuning on the cavity, the unitary may be reversed and a measurement on the non-interacting degree of freedom may be used to determine the phase. We simulate this scheme using a phase encoded on the atomic degree of freedom and a momentum measurement. To diagnose the metrological use, we consider the fidelity between the $\ket{+1}^{\otimes N}$ momentum state and the final momentum state. This is the same as measuring if $\langle\hat{P}_{COM}\rangle$ is equal to $+N \hbar k / 2$ or not. We consider this measurement outcome because for values of $\phi_1$ near zero, a $\hat{K}^z$ measurement outcome of $+N / 2$ is the most likely outcome, and for $\phi_1 = 0$, it will be the only outcome. As a result, this fidelity forms an effective probability distribution of $\phi_1$ for just this one measurement outcome of $\hat{K}^z$, and groups together the rest of the possible measurement outcomes. In Appendix~\ref{sec:FidExp} we show that this effective probability distribution provides a lower bound for the CFI. The standard deviation of this distribution may be used to calculate a lower bound for the CFI of this measurement scheme. The standard deviation of this fidelity is shown in Fig.~\ref{fig7} (a), while the inverted form of the standard deviation from equation Eq.~\eqref{eq:CRB_inv} is compared to the relevant QFIM diagonal element and shown in Fig.~\ref{fig7} (b). Using the fidelity to represent only one of the possible measurement outcomes, the uncertainty scales at $1/\sigma_{Fid}^2\approx 0.1699 N^{2}$ and from this we see that these auxiliary measurements allow us to predict the real phase with an uncertainty that scales with at least $59\%$ of the quantum CRB. This demonstrates that the auxiliary measurement, while not optimal compared to a direct measurement, still recovers a large amount of information about the degree of freedom not being directly measured.

\section{Conclusion}\label{sec:Conc}
In this work, we have introduced a novel method which individually squeezes and entangles two degrees of freedom, and showed there exists a non-trivial interplay between the atomic internal and momentum degrees of freedom. We have demonstrated that these extra degrees of freedom might create the opportunity for multi-parameter metrology at the Heisenberg limit in either degree of freedom, or for novel metrology schemes which benefit from the entangled degrees of freedom. The multiparamter and auxiliary schemes shown in the final section have potential to be the basis for practical tools in matter wave interferometry. This form of entanglement generation and manipulation represents a possible new frontier for atom interferometry.

Future work could include adding decoherence in a numerical exploration~\cite{ImprovedMixedQM_Luisa}, and explorations of the existence of multipartite entanglement\cite{mulitpartite_Ren} that may be realized by this system. We also note that the physical system explored here might pose experimental challenges. Namely, the regime requiring $\Delta \gg \sqrt{N} g$ leads to the parameter $\chi$ being small, thereby requiring long interaction times which are hard to achieve in atomic beam-cavity experiments. To explore the effects of the small $\chi$ and long interaction times compared to the decoherence time, one could simulate this system with full beam dynamics. It would also be interesting to explore the use of a moving optical lattice~\cite{Conveyor_Browaeys} to select the atomic transverse momentum, and trap the atoms in the cavity longer. We are also interested in the possibility of using the auxiliary measurement scheme for much shorter interaction times than shown here, $\chi t \ll 1$, such that the degrees of freedom only become weakly correlated and measurements on one degree of freedom only perturbatively affect the other degree of freedom. This could allow for measurements which only extract a small amount of information, but don't destroy the quantum state of the other degree of freedom.

Lastly, we point out that the above discussion is centered on realizing Eq.~\ref{eq:Hphys}, however the principles discussed here may be relevant to different platforms. Specifically, we believe coherently controlling a two-component Bose-Einstein
condensate~\cite{Spinor_Lev,SOC_Chen} in order to select for interactions, and engineering an optical lattice to induce spin-momentum couplings in a Bose-Einstein~\cite{spinMoment_Khamehchi} might lead to similar Lie algebraic structure and allow for controlled generation of metrologically useful entanglement. The use of a two component BEC might have the added benefit of relaxing the condition on small $\chi$ that we have here~\cite{SOC_Chen}.

\section*{Acknowledgments}
The authors thank John Cooper, Liang-Ying Chih, and Christopher Wilson for stimulating discussions.
This research was supported by the NSF PFC Grant No. 1734006 and the NSF Q-SEnSE Grant No. OMA 2016244. M.H. acknowledges support from a Visiting Fellowship from the University of Pisa, Italy. M. L. C. acknowledges support from the MIT-UNIPI program.
\bibliography{references.bib}

\pagebreak

\appendix

\section{Schwinger Boson Representation} \label{sec:SchwingerBoson_Operators}


\subsection{Normalization Coefficient} \label{sec:Normalization}

The symmeterizer in Eq.~\eqref{eq:SchwingState} is defined with the normalization factor, $1/\mathcal{M}_{\alpha,\beta,\beta,\delta}$, such that
\begin{equation}
    \mathcal{M}_{q,q_3,\sigma_3} = \sqrt{\frac{N!}{\alpha!\beta!\gamma!\delta!}},
\end{equation}
so that the bosonic state representation is normalized. In fact, we can see that $\mathcal{M}_{\alpha,\beta,\beta,\delta}^2$ is just a multinomial coefficient so this normalization makes our bosonic modes match a straightforward second quantization of the system's degrees of freedom.

\subsection{Creation and Annihilation Operators} \label{sec:create_annihilate}

For posterity, we present the remaining three annihilation operators not shown in the paper:
\begin{align}
    \hat{b} \ket{\alpha,\beta,\gamma,\delta} &= \sqrt{\beta} \ket{\alpha,\beta-1,\gamma,\delta}, \\
    \hat{c} \ket{\alpha,\beta,\gamma,\delta} &= \sqrt{\gamma} \ket{\alpha,\beta,\gamma-1,\delta}, \\
    \hat{d} \ket{\alpha,\beta,\gamma,\delta} &= \sqrt{\delta} \ket{\alpha,\beta,\gamma,\delta-1}.
\end{align}

\section{Interaction Picture for the Simplified Hamiltonian} \label{sec:InterHamil}

Starting with the Hamiltonian Eq.~\eqref{eq:HSU4}, we can choose a different interaction picture, such that
\begin{align}
    \hat{H} - \chi \hat{J}^z & = \chi ( \hat{E}^2 - (\hat{J}^z)^2 + \hat{J}^z ) - \hat{J}^z ) \nonumber \\
    & = \chi ( \hat{E}^2 - (\hat{J}^z)^2).
\end{align}
This is equivalent to choosing $\hat{H}_0 =  \sum_{j=1}^N \hbar \omega_a \hat{\sigma}^z_j / 2 + \chi \sum_{j=1}^N \hat{\sigma}^z_j / 2 + \hbar \omega_a \hat{a}_c^\dagger \hat{a}_c $ for our transformation into the interaction picture. This leads to an extra phase on the Pauli raising operator for the $j^{th}$ atom, so $\hat{\sigma}^+_j(t) = e^{(-i \chi t)} \hat{\sigma}^+_j$ in the interaction picture. However, this phase cancels after the adiabatic elimination of the cavity mode. Thus, we may effectively ignore the $\hbar \hat{J}^z$ appearing in the Hamiltonian. Our Hamiltonian is then
\begin{widetext}
    \begin{equation}
        \begin{aligned}
            \hat{H} &= \chi  \sum_{i,j=1}^N\hat{s}_{i}^x\hat{s}_{j}^x\hat{\sigma}_{i}^+\hat{\sigma}_{j}^- - \chi \sum_{j=1}^N \hat{\sigma}^z_j / 2 \\
            & = \chi ( \hat{a}^{\dagger} \hat{b} + \hat{c}^{\dagger} \hat{d} ) ( \hat{a} \hat{b}^{\dagger} + \hat{c} \hat{d}^{\dagger} ) - \frac{\chi}{2} ( \hat{a}^\dagger \hat{a} + \hat{c}^\dagger \hat{c} - \hat{b}^\dagger \hat{b} - \hat{d}^\dagger \hat{d} )\\
            & = \chi ( \hat{a}^{\dagger} \hat{a} \hat{b}^{\dagger}\hat{b} + \hat{c}^{\dagger}  \hat{c} \hat{d}^{\dagger}\hat{d} + \hat{a}^{\dagger} \hat{b} \hat{c} \hat{d}^{\dagger} + \hat{a} \hat{b}^{\dagger} \hat{c}^{\dagger} \hat{d} + \hat{a}^{\dagger} \hat{a} + \hat{c}^{\dagger}  \hat{c} - \frac{1}{2} \hat{n}_e + \frac{1}{2} \hat{n}_g ) \\
            & = \chi ( \hat{a}^{\dagger} \hat{a} \hat{b}^{\dagger}\hat{b} + \hat{c}^{\dagger}  \hat{c} \hat{d}^{\dagger}\hat{d} + \hat{a}^{\dagger} \hat{b} \hat{c} \hat{d}^{\dagger} + \hat{a} \hat{b}^{\dagger} \hat{c}^{\dagger} \hat{d} + \frac{1}{2} \hat{n}_e + \frac{1}{2} \hat{n}_g ) \\
            &= \chi ( \hat{E}^2 - (\hat{J}^z)^2 ).
        \end{aligned}
    \end{equation}
\end{widetext}

In the last line we have used the fact that $\frac{1}{2} \hat{n}_e + \frac{1}{2} \hat{n}_g = N/2$, and dropped this term due to it only contributing a global phase.

\section{Bayesian Inferencing Algorithm} \label{sec:Bayes}
In Section~\ref{sec:InterScheme}, we use Bayes theorem to carry out Bayesian inferencing. We aim to construct a probability distribution $P(\vec{\phi}|\vec{\epsilon})$, where $\vec{\phi} = (\phi_3,\phi_5)$ and $\vec{\epsilon}$ is a measurement log derived from a weighted random sampling of possible measurement outcomes. Here, we use the fact that $\hat{J}^x=\sum_i \lambda^x_i \hat{\Pi}^x_i$ and $\hat{K}^z=\sum_j \lambda^z_j \hat{\Pi}^z_j$, for eigenvalues $\lambda^x_i,\lambda^z_j$ and projective operators $\hat{\Pi}^x_i,\hat{\Pi}^z_j$. Both sets of projective operators form a complete positive operator valued measure on the set of states.

We simulate a measurement by choosing an outcome, $\epsilon_{i,j}$, corresponding to finding eigenvalue $\lambda^x_i$ for a $\hat{J}^x$ measurement and $\lambda^z_j$ for a $\hat{K}^z$ measurement. This outcome is chosen at random by sampling a list of all possible outcomes with weights $P(\epsilon_{i,j}) = \bra{\psi} \hat{V}^\dagger \hat{\Pi}^x_i \hat{\Pi}^z_j \hat{V} \ket{\psi} $, where $\hat{V}$ is given in Eq.~\eqref{eq:Signal} and $\ket{\psi} = \exp(-i t \hat{H})\ket{\psi_0}$. Through this process we generate the measurement log, $\vec{\epsilon}$.

We start with a flat prior distribution, $P(\vec{\phi})=(2\pi)^{-2}$, and update our probability distribution with each measurement outcome according to
\begin{equation}
    P_{m+1}(\vec{\phi}|\epsilon_{i,j}) = P(\epsilon_{i,j}|\vec{\phi})\frac{P_{m}(\vec{\phi})}{P(\epsilon_{i,j})},
\end{equation}
where $P(\epsilon_{i,j}|\vec{\phi})=\bra{\psi} \hat{V}_{\text{est}}^\dagger(\vec{\phi}) \hat{\Pi}^x_i \hat{\Pi}^z_j \hat{V}_{\text{est}}(\vec{\phi}) \ket{\psi}$ with $\hat{V}_{\text{est}}(\vec{\phi})$ being a numerical reconstruction of the unitary, $P(\epsilon_{i,j})$ is the probability of the measurement outcome integrated over all values of $\phi_3,\phi_5$, $P_{m}(\vec{\phi})$ is the probability distribution from the first $m$ measurements, and $P_{m+1}(\vec{\phi}|\vec{\epsilon})$ is the updated probability distribution.

We can predict the asymptotic behavior of Bayesian analysis from the classical Fisher information (CFI). The CFI can be explicitly calculated:
\begin{equation}
    I(\hat{G}^i) = \sum_i \left(\frac{d}{d \phi_i} \ln(P(\epsilon_j|\phi_i)) \right)^2 P(\epsilon_j|\phi_i) ,
\end{equation}

where $\epsilon_j$ represents the $j$\textsuperscript{th} measurement outcome, and $P(\epsilon_j|\phi_i)$ is the same probability distribution, but marginalized over any independent variables besides $\phi_i$. For example, if $i=3$ such that $\hat{G}^3=\hat{J}^z$, we have that

\begin{equation}
    P(\epsilon_j|\phi_3) = \Tr_J\left[ \hat{\Pi}^x_j \ \Tr_K( \hat{V}_{\text{est}}(\vec{\phi}) \ket{\psi}\bra{\psi} \hat{V}^\dagger(\vec{\phi}))\right],
\end{equation}

where $\Tr_J$ and $\Tr_K$ are the traces over the atomic internal degree of freedom, and momentum degree of freedom respectively. The CFI we consider is only dependent on a single degree of freedom because we only use it in a comparison to a diagonal element of the QFIM.

\section{Fidelity as a Lower Bound of the CRB} \label{sec:FidExp}

The trace of one degree of freedom in this system is very hard to caclulate, even just numerically, because correlations between the atomic energy level and momentum states happen on an atom by atom basis--whereas large simulations are only feasible using the 2nd quantization picture we show in this paper. This provided challenges for calculating the scaling behavior of the axuiliary scheme, where one wants to measure one degree of freedom but no the other. Here we briefly show that the fidelity between an eigenstate of an observable and the state one wishes to measure serves as a suitable lower bound on the actual set of measurements, without needing to take the trace of a degree of freedom. 

One may analytically calculate the CFI with respect to a $\hat{J}^x$ rotation and a full $\hat{K}^z$ measurement as follows,
    
\begin{equation}
    I(\hat{J}^x) = \sum_{m = -N/2}^{+N/2} p_j  \left( \frac{\partial}{\partial \phi_1} \log(p_j) \right)^2
\end{equation}
where $p_j$ represents the probability of the $j^\text{th}$ measurement outcome of $\hat{K}^z$, for example $p_{N/2} = \bra{+1}^N \text{Tr}_J( \ket{\psi}\bra{\psi} )\ket{+1}^N$, where $\text{Tr}_J$ means we first trace over the atomic degrees of freedom. We also have that $p_j  \left( \frac{\partial}{\partial \phi_1} \log(p_j) \right)^2 \geq 0$ for all outcomes $j$, so we can observe that
    
\begin{widetext}
\begin{equation}
    p_{N/2}  \left( \frac{\partial}{\partial \phi_1} \log(p_{N/2}) \right)^2 + p_{j\neq N/2}  \left( \frac{\partial}{\partial \phi_1} \log(p_{\neq N/2}) \right)^2 \leq \sum_{m = -N/2}^{+N/2} p_j  \left( \frac{\partial}{\partial \phi_1} \log(p_j) \right)^2,
\end{equation}
\end{widetext}
where $p_{j\neq N/2} = 1 - p_{N/2}$ is the probability of not measuring $\hat{K}^z = + N/2$. These two probabilities, $p_{N/2}$ and $p_{j\neq N/2}$ can be calculated without the use of a trace via the fidelity. This is because one may observe that under the time evolution of the Hamiltonian, the only atomic states entangled to $\ket{+1}^N$ momentum states are the ones in the initial atomic configuration, $\ket{+}^N$. Otherwise, momentum flips occur in pairs from the $\hat{s}^x_i \hat{s}^x_j$ term in the Hamiltonian. Therefore, the CFI of this single probability distribution, $p_{N/2}$, serves as a lower bound for the CFI by construction, because this would be the same as measuring if $\hat{K}^z$ is $+N/2$ or not.

\end{document}